\documentclass[12pt]{iopart}

\usepackage{graphicx}

\def	\pt        {p_{T}}

\def  \abseta    {\left|\eta\right|}
\def  \absetatrigger    {\left|\eta^{trig}\right|}
\def  \absetaassoc    {\left|\eta^{assoc}\right|}
\def  \deta    {\Delta\eta}

\def  \absdphi   {\left|\Delta\phi\right|}
\def  \s			   {\sqrt{s_{NN}}}

\begin{document}

\title[Jet-like correlations between Forward - and Mid - rapidity]{Jet-like correlations between Forward- and Mid- rapidity in p+p, d+Au and Au+Au collisions from STAR at $\s $= 200 GeV}

\author{Levente~Molnar (for the STAR Collaboration)}

\address{Purdue University, 525 Northwestern Avenue, West Lafayette, IN 47907-2036, USA;
INFN Sezione di Bari, Via E. Orabona 4, 70126 Bari, Italy}

\ead{Levente.Molnar@ba.infn.it}

\begin{abstract}
In this proceedings we present STAR measurements of two particle azimuthal correlations
between trigger particles at mid-rapidity ($\abseta<$ 1)  
and associated particles at forward rapidities (2.7 $<\abseta<$ 3.9)
in p+p, d+Au and Au+Au collisions at $\s $= 200 GeV. 
Two particle azimuthal correlations between a mid-rapidity 
trigger particle and forward-rapidity associated particles preferably 
probe large-x quarks scattered off small-x gluons in RHIC collisions. 
Comparison of the separate d- and Au-side measurements in d+Au collisions
may potentially probe gluon saturation and the presence of Color Glass 
Condensate.
In Au+Au collisions quark energy loss can be probed at large rapidities,
which may be different from gluon energy loss measured at mid-rapidity.
\end{abstract}


\section{Introduction}
Jet-like azimuthal correlations at mid-rapidity 
in Au+Au collisions at RHIC energies have shown significant modifications,
indicating the presence of a dense and strongly interacting medium~\cite{StarWhite}. 
Such measurements at mid-rapidity mainly probe the energy loss of 
gluons, due to the dominance of gluon-gluon scattering at RHIC energies~\cite{CTEQ}. 
STAR has the unique capability to extend two particle correlation measurements
to forward rapidities (2.7 $<\abseta<$ 3.9) by utilizing the Forward Time Projection Chambers (FTPC)~\cite{FtpcNim},
with trigger particles at mid-rapidity ($\abseta<$ 1) in the STAR-TPC~\cite{TpcNim}.
This kinematics is sensitive to hard scattering of small-x gluons 
(from which the trigger particles are mostly produced) and large-x quarks 
(which fragment into associated particles at forward rapidities). 
The measurements may address the question of gluon saturation in d+Au collisions, 
energy loss of quark jets at large rapidities and the possible presence of long 
range $\deta$ correlations in Au+Au collisions~\cite{Joern}.

\section{Analysis details}
High-$\pt$ trigger particles (3 $< \pt^{trig} <$ 10 GeV/c, $\absetatrigger<$ 1) 
and associated particles (0.2 $< \pt^{assoc} <$ 2 GeV/c, 2.7 $< \absetaassoc <$ 3.9) are selected in the TPC and FTPCs, respectively. 
The two particle correlation functions are corrected for tracking 
efficiency and acceptance of the associated particles, and are normalized 
per trigger particle. 
%
%
\begin{figure}[htb]   
\begin{center}
\vspace*{-0.3cm}
\resizebox{.5\textwidth}{!}{\includegraphics{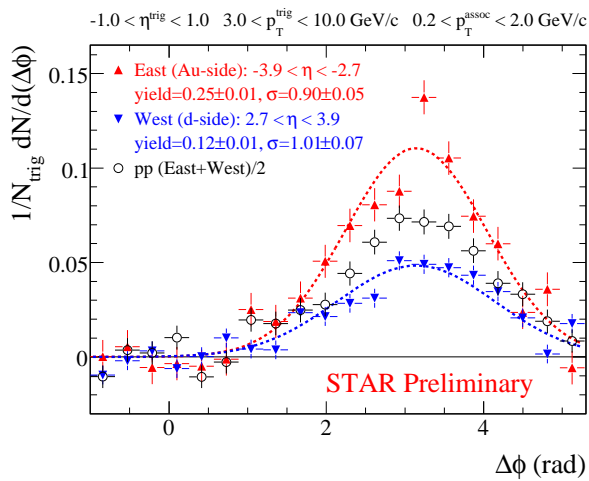}}
\vspace*{-0.1cm}
\caption[]{(color online) Correlation functions in d+Au collisions for the outgoing d-side (blue) and the Au-side (red) compared to pseudo-rapidity averaged p+p collisions (black) at $\s $= 200 GeV/c.}
\vspace*{-0.3cm}
\end{center}
\label{fig:ppdAu}
\end{figure}
%

The combinatorial background is constructed from the technique of event mixing.
In p+p and d+Au collisions, combinatorial background is normalized to the signal region of
$\absdphi<$ 1, because in this region no correlation structure is observed.
In Au+Au collisions, elliptic flow ($v_2$) modulation is added pairwise to the mixed-events background. 
The background is normalized to the range of $0.8<\absdphi<1.2$ by the Zero Yield At 1 (ZYA1) method~\cite{StarJet2,Ulery:2006iw}.
Trigger particle $v_2$ is taken to be the average of the results from the modified reaction plane and the 4-particle cumulant 
methods, and the range of the two results is taken to be the systematic uncertainty, as in~\cite{StarJet2}.
The associated particle $v_2$ used here is a parameterization of 2-particle cumulant measurements at forward rapidities by STAR~\cite{StarFtpcV2}, while the centrality dependence is parameterized from preliminary STAR results. 
The FTPC $v_{2}$ results obtained from different methods are in good agreement~\cite{StarFtpcV2}.   
However, as a conservative estimate the same relative systematic uncertainty as at mid-rapidity is applied.
%
%
%
\begin{figure}[htb]   
\begin{center}
\resizebox{.45\textwidth}{!}{\includegraphics{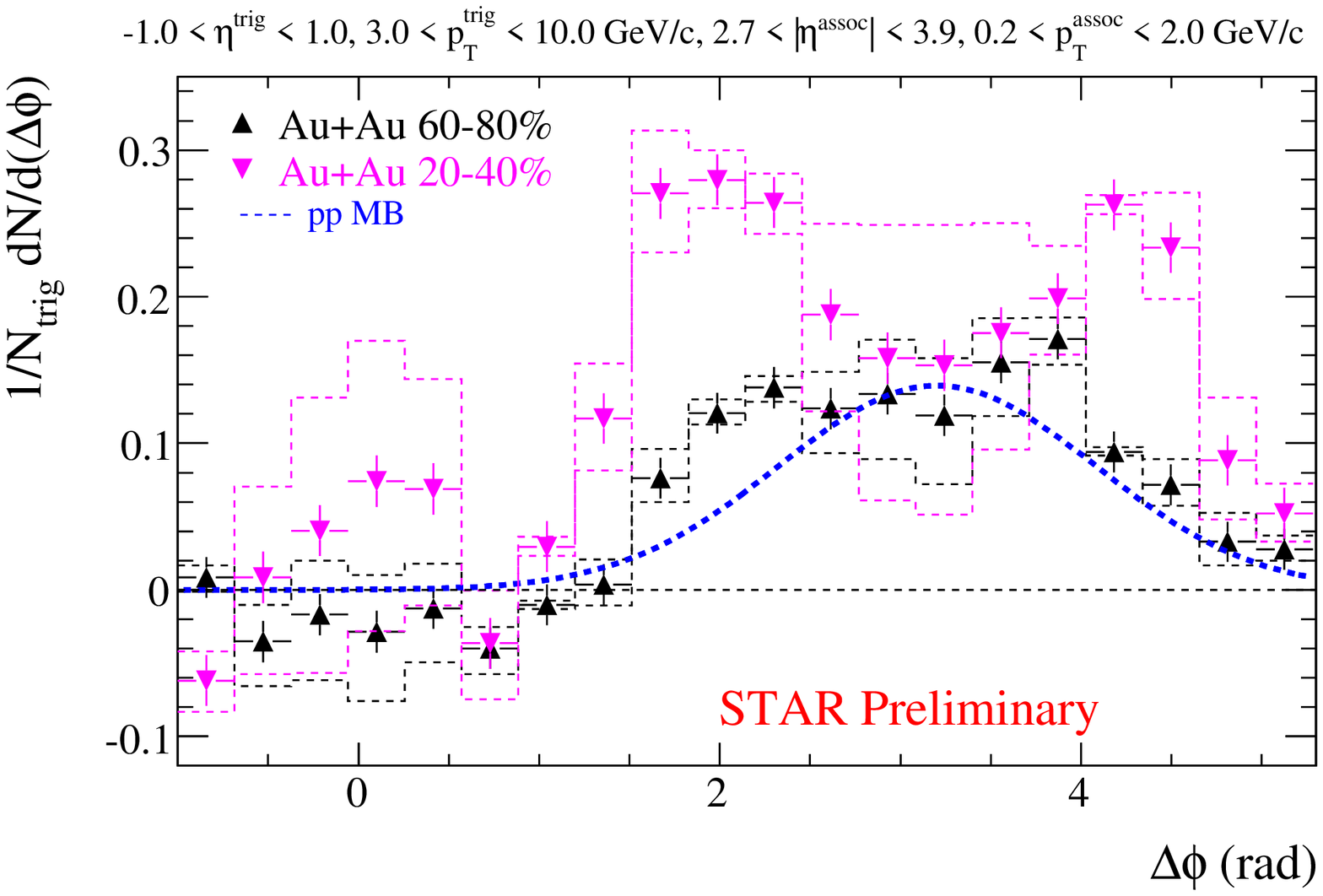}}
\resizebox{.45\textwidth}{!}{\includegraphics{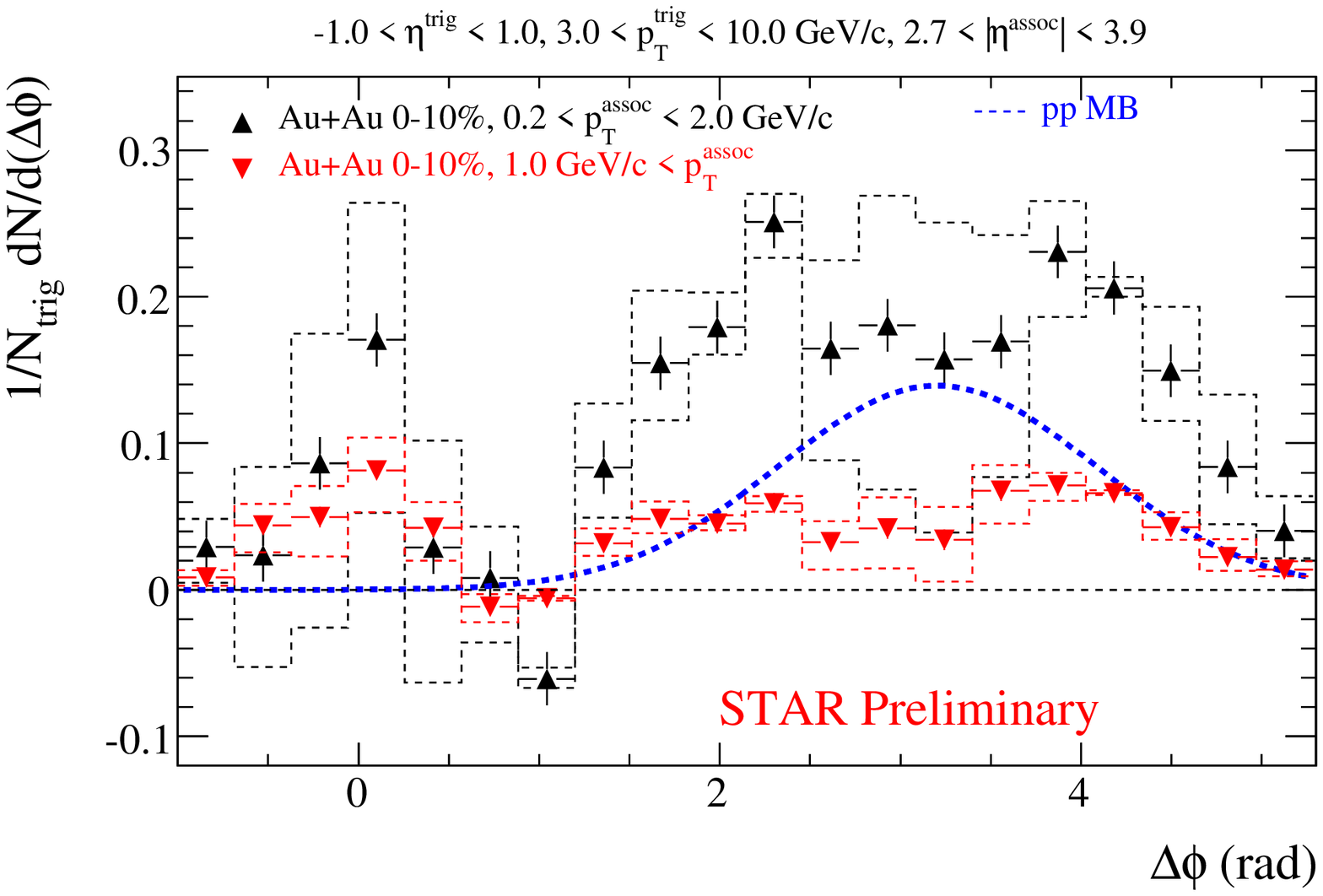}}
\vspace*{-0.1cm}
\caption[]{(color online) Left panel: correlation functions in 60-80\% and 20-40\% Au+Au collisions at 200 GeV/c. Right panel: correlation functions for two associated momentum ranges in 0-10\% Au+Au collisions at 200 GeV/c. Curve represents a fit to the correlation function in p+p collisions at $\s $= 200 GeV/c.}
\vspace*{-0.3cm}
\end{center}
\label{fig:AuAu}
\end{figure}
%
%
%
\section{Results and discussion}
\subsection{d+Au results - probing small-x gluons in nucleus}
Figure 1 shows the two particle azimuthal correlations for 
p+p and d+Au collisions at $\s$ = 200 GeV. 
The outgoing d- and Au-sides are shown separately. 
The p+p points are averaged over the positive and negative rapidities. 
As shown in Fig. 1, the away-side correlation shapes are similar. The 
d-side yield is suppressed by about a factor of two compared to the Au-side. 
The p+p result lies approximately at the average in between.
%
%
%
\begin{figure}[b]   
\begin{center}
\resizebox{.30\textwidth}{!}{\includegraphics{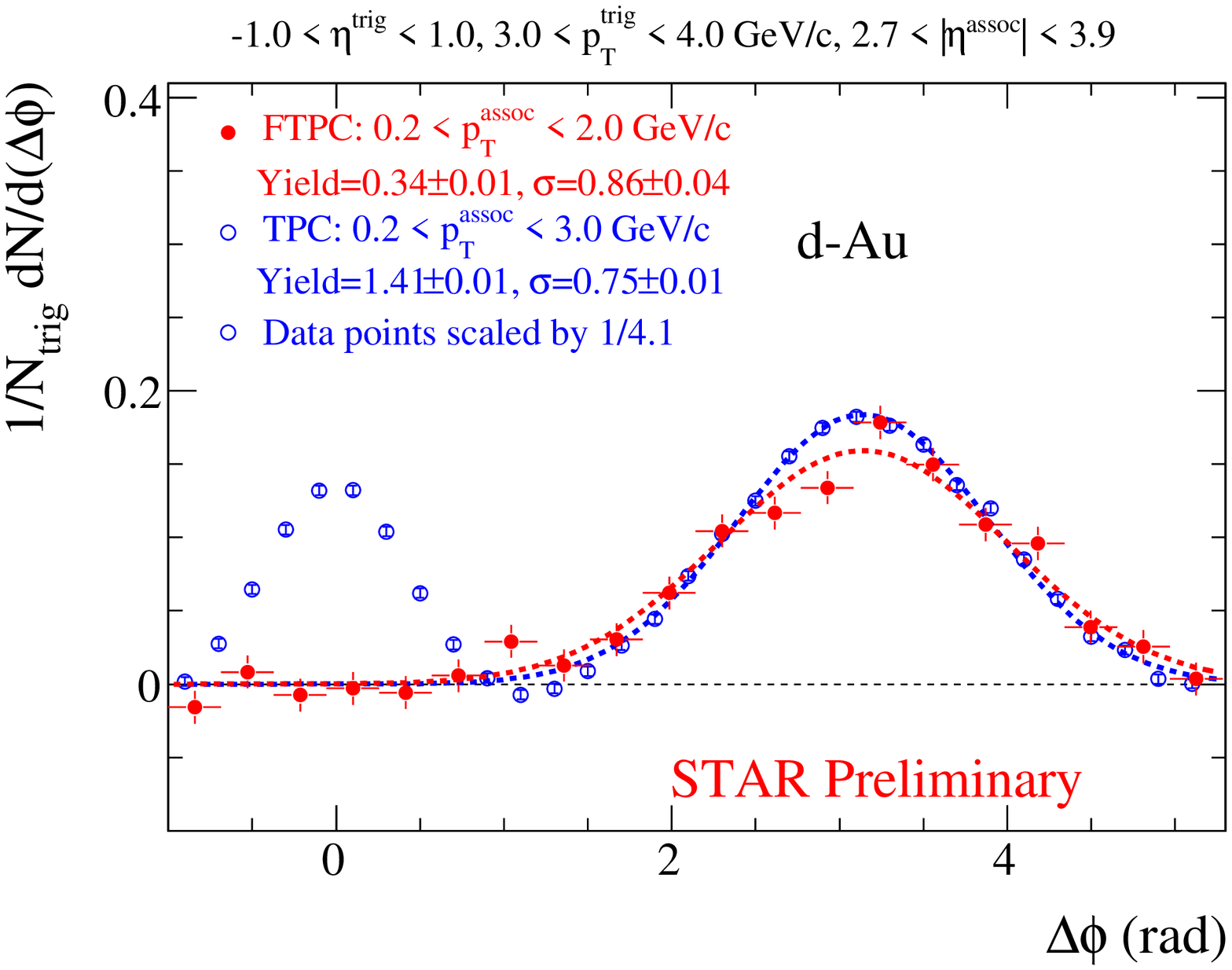}}
\resizebox{.30\textwidth}{!}{\includegraphics{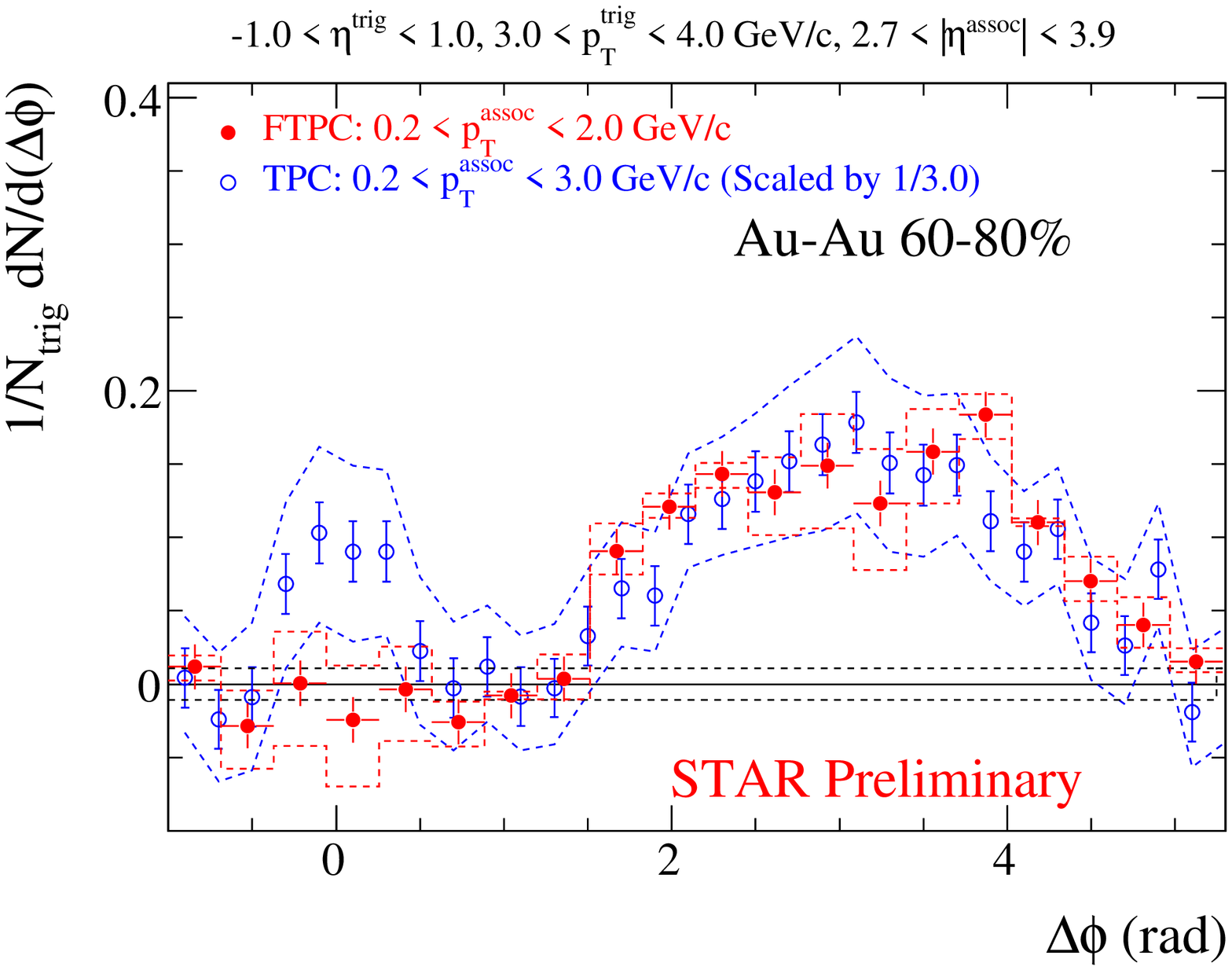}}
\resizebox{.30\textwidth}{!}{\includegraphics{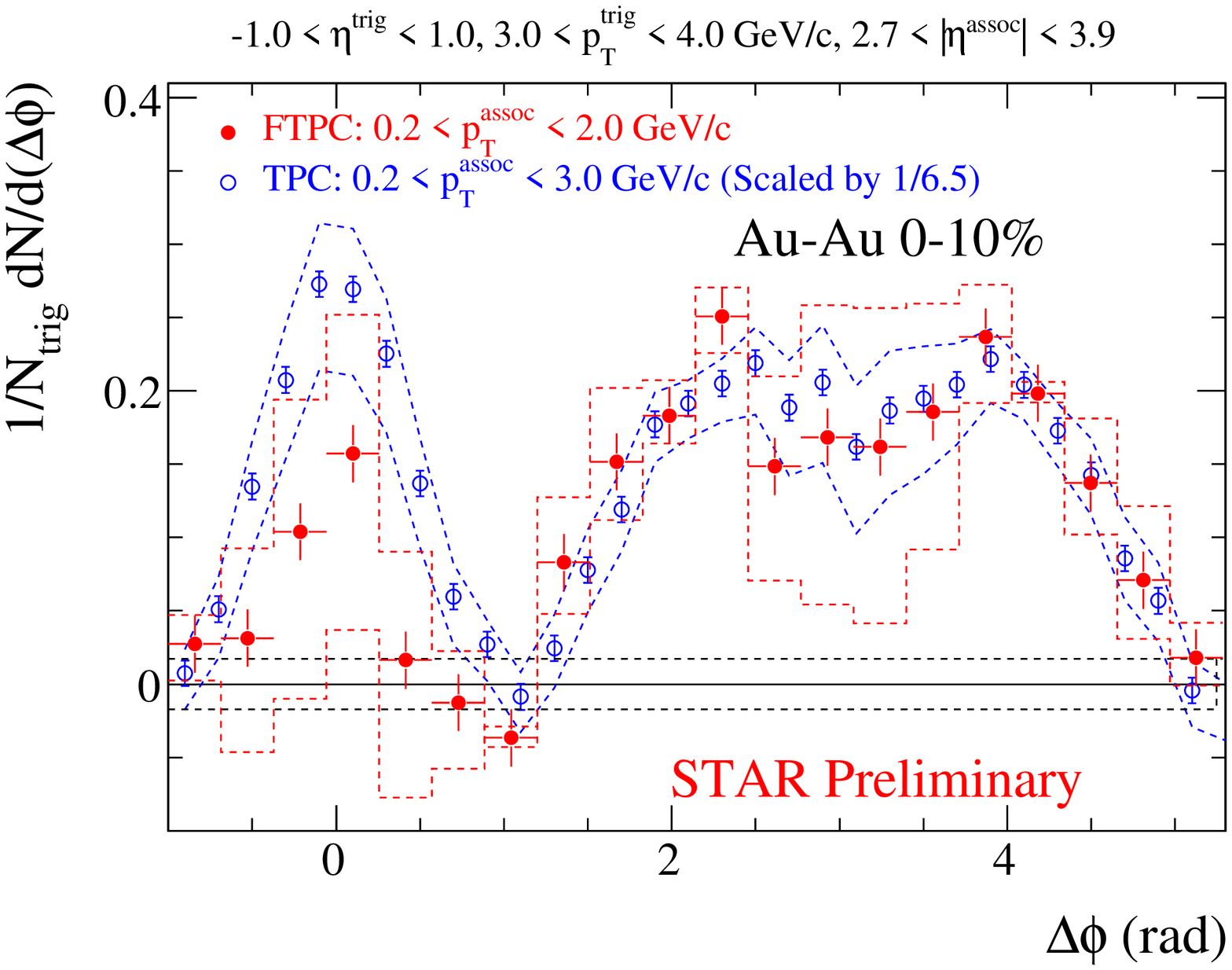}}

\vspace*{-0.1cm}
\caption[]{(color online) Comparison of correlation functions from TPC (blue) and FTPC (red) for d+Au, peripheral (60-80\%) and central (0-10\%) Au+Au collisions at $\s $= 200 GeV/c. The TPC data are scaled to compare the correlation shapes. Dashed lines represent systematic uncertainties of the Au+Au data.}
\vspace*{-0.3cm}
\end{center}
\label{fig:TPCFTPC}
\end{figure}
%
%
Suppression of the d-side yield may be understood as the 
suppression of small-x gluons in the Au nucleus, as predicted by the 
Color Glass Condensate (CGC) picture~\cite{CGC}. On the other hand, reduction of the d-side yield may 
arise from the energy degradation of the d-side quarks due to multiple scattering in the Au 
nucleus. Gluon anti-shadowing~\cite{antishadowing} and the EMC effect~\cite{EMC}
would enhance the d-side yield relative to the Au-side.
\subsection{Au+Au results - rapidity dependence of energy loss}
Figure~\ref{fig:AuAu} shows the azimuthal correlations for 60-80\% and 20-40\% (left panel) and for 0-10\% (right panel) Au+Au collisions at $\s$ = 200 GeV. Results for two associated $\pt$ ranges are shown for the 0-10\% Au+Au data.
The near-side correlations for 0.2 $< \pt^{assoc} <$ 2 GeV/c are consistent with zero within the systematic uncertainties.
However, the central data at high associated $\pt$, with the reduced systematic uncertainty, are suggestive of non-zero correlation on the near-side. This result indicates that long range $\deta$ correlations, first observed in $\deta<$ 2~\cite{Joern}, may extend out to $\deta \sim$ 4 in the FTPCs.

The away-side correlation shapes in Au+Au collisions are broadened with respect to p+p as shown in Fig.~\ref{fig:AuAu}. 
The broadening is present for each centrality and is similar to mid-rapidity measurements. 
Figure 3 shows the comparison of the azimuthal correlations measured at forward rapidity (red) and at mid-rapidity (blue). The away-side correlation shapes are identical within the systematic uncertainties.
To understand the contributing physical processes of energy loss at
mid- and forward-rapidities quantitative theory calculations are essential.

\section{Summary}
In summary, we have presented two particle azimuthal correlations 
of charged hadrons at forward rapidities (2.7 $< \abseta <$ 3.9) 
with trigger particles selected at mid-rapidity ($\abseta <$ 1) 
in p+p, d+Au and Au+Au collisions at $\sqrt{s_{NN}}=$ 200 GeV. 
\\
\\
Near-side correlation is not observed in p+p and d+Au collisions.
Near-side correlations in Au+Au collisions are consistent with zero with the systematic uncertainties for 0.2$<\pt^{assoc}<$2 GeV/c. 
However, the high associated $\pt$ data ($\pt>$1 GeV/c) in central Au+Au collisions suggest a finite near-side correlation, which may
indicate the presence of long range $\deta$ correlations at forward rapidities.
\\
\\
Significant away-side correlations are observed for all systems. 
In d+Au collisions, a factor of two suppression is observed for the d-side compared to the Au-side. The p+p measurement is approximately the average of the d-side and Au-side correlations.
While the Color Glass Condensate gives a qualitative description of the observed relative d- and Au-side yields, 
due to suppression of small-x gluons in Au nucleus, other mechanisms may be also at work, such as $d$ energy degrading.
Quantitative model calculations are needed for further understanding.
The away-side correlation shape broadens from peripheral to central Au+Au collisions, 
similar to mid-rapidity observations. Moreover, the away-side correlation 
shapes are almost identical to those measured at mid-rapidity. 
To understand the contributing physical processes of energy loss at
mid- and forward-rapidities quantitative theory calculations are essential.

\end{document}